\documentstyle[12pt]{article}
\def\beq{\begin{equation}}
\def\eeq{\end{equation}}
\def\bea{\begin{eqnarray}}
\def\eea{\end{eqnarray}}
\def\dsp{\displaystyle}

\begin{document}

\newcommand{\sheptitle}
{Effective Potential and KK-Renormalization Scheme in a
5D Supersymmetric Theory}

\newcommand{\shepauthor}
{Vicente Di Clemente  and  Yuri A. Kubyshin\footnote{Permanent address:
Institute of Nuclear Physics, Moscow State University, Moscow 119899, 
Russia}}

\newcommand{\shepaddress}
{Department of Physics and Astronomy,
University of Southampton, Southampton, SO17 1BJ, U.K.}

\newcommand{\shepabstract}
{We calculate the 1-loop effective potential
in a supersymmetric model in 5D with $S^1/(Z_2\times Z_2)$
orbifold compactification. The procedure of calculation consists of
evaluating first the integrals over four-momenta using the
dimensional regularization and then the sum over Kaluza-Klein
modes using the zeta-regularization. We show that both fermionic
and bosonic contributions are separately finite and
argue that, supersymmetry is not necessary for the finiteness
of the theory at 1-loop. Also, some general arguments on the
finiteness of the theory with arbitrary number
of extra dimensions are presented.}

\begin{titlepage}
\begin{flushright}
hep-th/0108117\\
SHEP 01-19
\end{flushright}
\vspace{0.5in}
\begin{center}
{\Large{\bf \sheptitle}}
\vspace{0.5in}
\bigskip \\ \shepauthor \\ \mbox{} \\ {\it \shepaddress} \\
\vspace{0.5in}
{\bf Abstract} \bigskip \end{center} \setcounter{page}{0}
\shepabstract
\end{titlepage}

%%%%%%%%%%%%%%%%%%%%%%%%%%%%%%%%%%%%%%%%%%%%%%%%%%%%%%%%%%%%%%%%%%%%%%%%%%%%%

\section{Introduction}

The Standard Model (SM) of electroweak and strong interactions,
providing a description of the elementary particles and three
of the fundamental interactions, has been rigorously tested at
high energy colliders with an excellent agreement.
However, there are reasons to believe that the SM is not a
complete theory. One of the fundamental problems of the SM is to
explain the origin of the electroweak symmetry breaking (EWSB) that 
leads to the known pattern of vector boson and fermion masses.
The only known perturbative mechanism of the EWSB is the Higgs mechanism
with the Higgs boson acquiring a vacuum expectation value.
As it is well known, in the SM the radiative corrections to the Higgs
boson mass are dominated
by ultraviolet quadratic divergences. As a consequence, were  
the SM a fundamental theory up to the Planck scale, the tree
level parameters of the theory would have to be fine tuned with a high
precision (fine tunning problem). Contrary to this, if the SM is a part 
of a supersymmetric theory with a low energy (TeV scale) supersymmetry 
(SUSY) breaking mechanism the divergence is logarithmic and the fine tuning
problem can be relaxed.

Much work has been done on five-dimensional (5D) generalizations
of the SUSY SM with the extra dimension being compact and of
large size $R \sim 1 /\mbox{TeV}^{-1}$~\cite{antoniadis} -
\cite{barbieri2}.
The main features of models of this kind are the following:
(i) they provide a mechanism of SUSY breaking;
(ii) the contribution of Kaluza-Klein (KK) modes to the mass term in the
efective potential is negative, thus triggering the EWSB via radiative
corrections;
(iii) the 1-loop radiative correction to the effective potential
is finite.

Such 5D theories may emerge as low energy limits of
string theories~\cite{antoniadis1}, in which case they have a natural
cutoff $\Lambda$ which is of the order of the string scale.
Because of their amasing softness the EWSB is not sensitive 
(at least at 1-loop
order~\footnote{Recently the 2-loop Higgs two-point function at zero external 
momenta was calculated and was shown
to be finite~\cite{mariano}; some general arguments on the absence
of ultraviolet divergencies at all loops in such theories were presented in
Ref.~\cite{masiero}.})
to $\Lambda$, i.e. the Higgs physycs is independent of the physics
at higher energies. In the present paper we will focus precisely
on the finiteness of the 1-loop contribution to the effective potential.

Let us remind that the 1-loop Higgs effective potential in a
5D generalization of the SM is equal to the infinite
sum of 1-loop contributions of KK modes, each of the contributions
being given by an integral over four-momenta. The index $k$, labelling
the KK modes, is essentially the discrete momentum along the fifth
direction (in units of $R^{-1}$). The contribution of an
individual KK mode is ultraviolet divergent, therefore, in general, 
the integration and summation do not commute. 
In a number of articles the 1-loop
effective potential was obtained by, firstly, calculating the sum
over KK modes, and then integrating the result over the four-dimensional
momentum \cite{delgado, barbieri1}. Such procedure of calculation
is often referred to as the KK-regularization. 
The validity of the calculation was questioned in some papers (see
Ref.~\cite{nieles, kim}), the main point of the discussion being 
the apparent non-commutativity of the integration and summation within 
this regularization. It was also argued
there that from the physical point of view it did not make sense 
to take into account the
whole infinite tower of KK modes and that rather only a finite number
of modes with masses below some effective cutoff scale $\Lambda$
should be included into the sum. Indeed, since a 5D theory is in general
nonrenormalizable, it is natural to think that there
exists a cutoff $\Lambda$ such that for energies above this
cutoff the physics is described by a more fundamental theory.
For instance, in Ref.~\cite{nieles} explicit cutoffs both in the KK
sum and in the momentum integral were introduced, and it was shown
that the finiteness of the result depends on whether the sum is taken
over a finite or infinite number of KK modes.
In fact, it is clear that cutting the sum explicitly breaks the
5D structure of the theory, so that it is not surprising to find
different ultraviolet behaviors for different ways of dealing with  
the divergencies. 
Alternatively, in Ref.~\cite{gero} both a momentum cutoff and an
appropriate symmetry preserving KK cutoff are used. Within this
framework contributions of high KK modes are exponentially suppressed and a
consistent finite result is obtained.

For regularizations which
do not violate essential symmetries of the theory, the result
after renormalization should not depend on the order of integration
and summation, as well as on the particular regularization used.
Among such ``good'' regularizations some are more convenient from the
practical point of view. In the present article we propose both 
a regularization of this kind and a renormalization scheme related to 
it. Our approach turns out to be quite effective, at least at 
1-loop order, and provides a deeper understanding of the origin 
of the softness of the theory.

As a concrete example we study the 1-loop contribution to the
effective potential in a 5D SUSY theory formulated in Ref. \cite{barbieri1}.
Our approach consists in calculating
first the four-dimensional integrals using the dimensional regularization,
and then evaluating the sum over the KK modes, using the $\zeta$-function
regularization \cite{zeta}. The idea to calculte the sums over KK modes in 
multidimensional theories by means of the $\zeta$-function 
regularization method is not new and was used in a number of 
papers (see for example \cite{ADPQ, zeta-KK, EliKub, corneliu}).  
Divergencies associated with the momentum
integration appear as simple poles in $2-n/2=\epsilon$. Here  
$n=4-2\epsilon$ is the dimension of the momentum space in 
regularized integrals, and $n \rightarrow 4$ when the regularization 
is removed. The $1/\epsilon$-pole in the 1-loop contribution of
a multidimensional field $X$ is multiplied by the sum of three terms:
(1) the term $\propto m_{X,k=0}^{4}$, where $m_{X,k=0}$
is the mass of the zero mode of $X$;
(2) a contribution of massive modes labelled by $k>0$
(we denote this set of modes by $X^{+}$);
(3) a contribution of massive modes with $k<0$
(we denote this set of modes by $X^{-}$).
We will show that the sum of these three terms is exactly zero
provided that the contributions of the {\it whole tower} of KK modes
are included.

Because of the finiteness of the 1-loop effective potential,
in principle no further subtraction of ultraviolet divergences is needed. 
However, when using the result without subtraction one should
bear in mind that a natural renormalization scheme, 
associated with the dimensional regularization of integrals and
$\zeta$-regularization of the sum, is implicitely assumed. 
We will refer to this renormalization scheme as the "KK-renormalization". 
The corresponding renormalization procedure consists basically
in subtracting poles in $\epsilon$. In 
the model, considered here, no subtraction
is actually needed because the coefficient
of the pole $1/\epsilon$ turns out to be zero. The reason of this
feature is the cancellation between the contributions of the 
three terms described above, namely the term $\propto m_{X,k=0}^{4}$, 
$X^{+}$ and $X^{-}$.

Important properties of the KK-renormalization
scheme are the following:
\begin{enumerate}
\item The bosonic and fermionic contributions to the 1-loop
effective potential are finite on their own, and the cancellation
between bosons and fermions due to SUSY in principle is not
needed.
\item No special fine-tuning of parameters is needed; this is in 
contrast with the case of the KK-regularization (see \cite{nieles}).
\item An important condition of the ultraviolet finiteness at 1-loop order
is that the spectrum of the massive modes taken into account in the sum 
must be the complete KK tower, from $k = -\infty$ to $k=\infty$.
\item Finally, the KK-renormalization scheme respects all the symmetry 
of the theory; in the case under consideration it is the 
${\cal N} = 2$ supersymmetry.
\end{enumerate}

Our result confirms the one of Refs. \cite{delgado, barbieri1}.

The structure of the paper is the following. In Sect. 2 we describe
the model and give a brief review of previous calculations of the 1-loop
effective potential. In Sect. 3 we calculate the potential using
the dimensional and $\zeta$-regularizations.
In Sect. 4 the limit of the zero size of the space of extra dimension 
is analyzed. 
Sect. 4 contains a discussion of the result and concluding remarks.

%%%%%%%%%%%%%%%%%%%%%%% Sect.2 THE MODEL %%%%%%%%%%%%%%%%%%%%%%%%%%%%%%%%%%

\section{The model}

We study a 5D SUSY extension of the SM formulated in
Ref.~\cite{barbieri1} (see also ~\cite{nomura}).
The extra dimension is compactified on the orbifold
$S^1/(Z_2\times Z_2^\prime)$ with the radius of the
circle being equal to $R$. The orbifold is constructed
by making two $Z_{2}$-indentifications of points $y$ of the
circle as follows: (1) $Z_2$ : $y\to -y$, and
(2) $Z_2^\prime$ : $y-\pi R/2 \to -y+ \pi R/2$.
All fields of the theory propagate in the bulk.
Brane interactions are located at $y=0$, $\pi R$ and $\pm \pi R /2$.

The SM matter and Higgs fields are described by a set of
${\cal N} =2$ hypermultiplets
$(\phi,\hat{\phi},\Psi)_X$, where $\Psi_X$ is a Dirac spinor with
components $(\psi,{\psi^{c}}^\dag)_X$;
$\phi_X$ and $\hat{\phi}_X$ are two complex scalars and $X$ runs over
three sets of
multiplets $Q,U,D,L,E$, corresponding to the three generations of matter, 
and a single Higgs $H$.
The gauge fields belong to the vector supermultiplet   
${\cal V}$ in the adjoint representation of the gauge group
$SU(3)\times SU(2)_L\times U(1)_Y$.
The on-shell field content of ${\cal V}$ is given by
${\cal V}=(V_M,\lambda^i,\Sigma)$, where $M=0,1,2,3,5$,
$\Sigma$ is a real scalar
field and $\lambda^i$, $i=\{1,2\}$ are two gauginos.
The latter reflects the presence of two supersymmetries in the theory 
obtained by the dimensional reduction to four dimensions. 
Namely, the hypermultiplets decompose into two 
four-dimensional ${\cal N} = 1$ chiral multiplets 
$X= (\phi_X,\psi_X)$ and $X^c = (\phi^c_X,\psi_X^c)$,
where $\hat{\phi} = {\phi^c}^\dag$. The $5D$ vector supermultiplets
can be decomposed into the four-dimensional vector supermultiplet
$V= (V_\mu,\lambda^1)$ ($\mu = 0,1,2,3$) and the chiral multiplet
$((\Sigma+iV_5), \lambda^2)$ in the adjoint representation of
the gauge group. In accordance with the discrete
$Z_2\times Z_2^\prime$-symmetry all fields in the theory have
one of the following $(Z_{2},Z_{2}')$ quantum numbers (parities):
$(+,+)$, $(+,-)$, $(-,+)$ or $(-,-)$.

Once both orbifoldings are imposed only the SM fields
have the $(+,+)$ parity and, hence, the zero modes.
The rest of the fields form infinite KK towers without
zero modes. The Yukawa interactions are introduced only 
at the fixed points of
the orbifold, i.e. at $y=0,\pi R, \pi R/2$ and $-\pi R/2$.
At low energies the effective theory is precisely the SM with
the tree-level Higgs potential proportional to
$|\phi_{H}|^{4}$ \cite{barbieri1}.

It is well known that the dominant radiative correction
to the Higgs potential comes from the top and stop KK towers whereas
the first and second generations and the gauge interactions
give relatively negligible contributions.
Since the initial theory is five-dimensional, all massive modes
circulate inside 1-loop Feynman diagrams.
The complete (all-orders) 1-loop effective potential is equal to
\beq
V_{\mbox{\footnotesize 1-loop}} = \frac{1}{2} Tr \left[ \sum_{k=-\infty}^{\infty} 
\int \frac{d^{4}p}{(2\pi)^{4}} \ln \left( 
\frac{p^{2} + (m^{(B)}_{k} (H))^{2}}{p^{2} + (m^{(F)}_{k} (H))^{2}} 
\right) \right],
\label{Veff}
\eeq
where $H = <\phi_{H}>$ is the Higgs background field, 
\beq
m^{(B)}_{k} (H) = \left|\frac{\pm 2k+1}{R} + m_{t} (H) \right|
 \label{mb}   
\eeq
and
\beq
m^{(F)}_{k} (H) = \left| \frac{\pm 2k}{R} + m_{t} (H)  \right|
\label{mf}
\eeq
($k=0,1,2, \ldots$) are the mass eigenvalues of the bosonic 
(top squark) and fermionic (top quark) KK states respectively,
\beq
m_{t}(H) = \frac{2}{\pi R} \arctan \frac{\pi y_{t} RH}{2}
\label{mtop} 
\eeq
is the mass of the zero mode top quark, and $y_{t}$ is the
four-dimensional Yukawa coupling.
The trace is taken over the degrees of freedom of the top
hypermultiplet for a given $k$, thus giving the factor
$N = 12$. 

Eqs (\ref{mb}) and (\ref{mf}) describe three classes
of mass eigenvalues for both bosons and fermions,
namely the zero mode with  $k=0$, $k>0$ modes and $k<0$ modes.
Note that though the compactification is on the orbifold the sum
over the KK modes goes from $k=-\infty$ to $k=\infty$. What happens is 
that after the orbifold compactification the net number of towers 
is divided by two but the positive non-zero modes of $X$
replace the negative non-zero modes of ${X^{c}}^{\dag}$, forming a
single tower with $k$ running from
$k=-\infty$ to $k=\infty$. This turns out to be a rather generic
property which takes place in a number of models (see Refs.~\cite{ADPQ},
\cite{benakli2}).

As it was already mentioned in the Introduction,
one of the procedures to obtain $V_{\mbox{\small 1-loop}}(H)$,
Eq. (\ref{Veff}), is to calculate first the sum and
then the integral. Evaluating each term of the sum we get
\begin{eqnarray}
V_{\mbox{\footnotesize 1-loop}} & \propto & \sum_{k=-\infty}^{\infty}
\int^{\infty} d\rho \rho \ln
\left(\frac{1+(m^{(B)}_{k} (H))^{2}/\rho}
{1+(m^{(F)}_{k} (H))^{2}/\rho}\right) \nonumber \\
& = & \sum_{k=-\infty}^{\infty}\int^{\infty} d\rho \left[ (m^{(B)}_{k}
(H))^{2} - (m^{(F)}_{k} (H))^{2} + {\cal O}\left(\frac{1}{\rho}\right)
\right], \label{sum-int}
\end{eqnarray}
where $\rho = |p|^{2}$. It is easy to see that
the contribution of each mode is ultraviolet divergent.
Because of this, in general, interchanging the summation and
integration in Eq. (\ref{Veff}) is not a well defined operation. 

If one first sums over the modes and then integrate over the four-momentum
the 1-loop effective potential becomes 
\begin{equation}
V_{\mbox{\footnotesize 1-loop}} =
N\int \frac{d^4 p}{(2\pi)^4} \left(W_B - W_F \right)
\label{int-sum}
\end{equation}
where
\begin{equation}
W_{B(F)} =  \frac{1}{2}\sum_{k=-\infty}^{\infty} \ln 
\left(R^2 \left[ p^2 + (m^{B (F)}_{k} (H))^{2} \right] \right)
\end{equation}
To perform summation one can calculate $dW_{B(F)}/d(R |p|)$
(the tadpole contribution) using the method of residues
and then integrate with respect to $R |p|$~\cite{delgado}. The result is
\begin{eqnarray}
W_B &=& \frac{1}{2}\left[ \pi R |p| + \ln\left(1+ r e^{-\pi R |p|}\right) +
\ln \left(1+ \frac{1}{r} e^{-\pi R |p|}\right) \right] \label{wb}\\
 W_F &=& \frac{1}{2}\left[ \pi R |p| + \ln \left(1- r e^{-\pi R |p|}\right)
 + \ln \left(1- \frac{1}{r} e^{-\pi R |p|}\right) \right], \label{wf}
\end{eqnarray}
were $r=\exp(i\pi R m_t(H))$.
The momentum integral of the first term in Eqs. (\ref{wb}), (\ref{wf}),  
the one $\propto \pi R |p|$, is divergent, but the divergencies 
cancel each other in the final expression, Eq. (\ref{int-sum}). 
It is easy to see that the rest of the terms in Eqs. (\ref{wb}), 
(\ref{wf}) give finite contributions to the 1-loop potential.
We would like to note that if the integral in Eq. (\ref{int-sum})
is calculated using the dimensional regularization, then
integration of the term $\propto \pi R |p|$ gives zero,
and $W_{B}$ and $W_{F}$ are separately finite.

Expanding (\ref{wb}) and (\ref{wf}) in powers of $r$ and
integrating over the four-momenta one obtains the result
of Ref. \cite{barbieri1}:
\begin{equation}
V_{\mbox{\footnotesize 1-loop}}(H) = \frac{3N}{2\pi^6 R^4}
\sum_{k=0}^{\infty}\frac{\cos\left[(2k+1)\pi R m_t(H)\right]}{(2k+1)^5}.  
\label{final}
\end{equation}

Note that in the above calculation the effective potential is
finite (for any regularization which does not break SUSY)
because the divergencies in the
bosonic and fermionic contributions cancel each other exactly due to SUSY.

If the regularization is consistent with symmetries of the theory, then
interchanging of summation and integration in Eq. (\ref{Veff})
should not alter the result.
In the next section we will use a regularization of this kind, namely 
a combination of the dimensional and $\zeta$-regularizations,
to calculate $V_{\mbox{\footnotesize 1-loop}}(H)$. We will first
perform the integration in Eq. (\ref{Veff}) and then the summation.
It turns out that with such regularization
the bosonic and fermionic contributions are separately finite,
and the final result for the 1-loop effective potential is in agreement 
with (\ref{final}).

%%%%%%%%%%%%%%%%%%%%%%% Sect.3 KK-RENORMALIZATION

\section{KK-Renormalization}

Let us write expression (\ref{Veff}) for the 1-loop term  
of the bare effective potential as the difference
of bosonic and fermionic contributions, 
\beq
V_{\mbox{\footnotesize 1-loop}}^{(bare)}(H) = 
V_{B}(H) - V_{F}(H),  \label{Veff-dif}
\eeq
and analyse first the bosonic contribution
\beq
V_{B}(H) = \frac{N}{2} \sum_{k=-\infty}^{\infty} 
\int \frac{d^{4}p}{(2\pi)^{4}} \ln \left( R^{2} \left[ 
p^{2} + (m^{(B)}_{k} (H))^{2} \right] \right). 
\eeq
Introducing the dimensional regularization one gets 
\bea
V_{B}(H) & = & \frac{N}{2} (\mu^{2})^{2-n/2}\sum_{k=-\infty}^{\infty} 
\int \frac{d^{n}p}{(2\pi)^{n}} \ln \left( R^{2} \left[ 
p^{2} + (m^{(B)}_{k} (H))^{2} \right] \right) \nonumber \\
& = & - \frac{N}{2} (\mu^{2})^{2-n/2} \frac{\partial}{\partial \alpha} 
\left. \sum_{k=-\infty}^{\infty} \int \frac{d^{n}p}{(2\pi)^{n}} 
\frac{1}{[R^{2}(p^{2} + (m^{(B)}_{k} (H))^{2}]^{\alpha}} \right|_{\alpha=0}, 
\label{Veff1}
\eea 
where $n=4-2\epsilon$ and $\mu$ is a mass scale which appears
due to the dimensional transmutation.
Using standard formulas we integrate over
momentum and obtain that
\bea
V_{B}(H) & = & - \frac{N}{2} \frac{(\mu^{2})^{2-n/2}}{(4\pi)^{n/2}} 
\frac{\partial}{\partial \alpha} \left. 
\frac{\Gamma (\alpha-n/2)}{4^{\alpha}R^{2\alpha} \Gamma (\alpha)} 
\sum_{k=-\infty}^{\infty} 
(m^{(B)}_{k}(H))^{n-2\alpha}\right|_{\alpha=0} \nonumber \\
& = & - \frac{N}{2} \frac{(\mu^{2})^{\epsilon}(4\pi)^{\epsilon}}{16\pi^{2}} 
\frac{\partial}{\partial \alpha} \left. 
\frac{ \Gamma (\alpha - 2 + \epsilon)}{ 4^{\alpha} R^{2\alpha} 
\Gamma (\alpha)} \sum_{k=-\infty}^{\infty} 
\frac{1}{\left[ \frac{2k+1}{R} + m_{t}(H) \right]^{2(\alpha-2+\epsilon)}}
\right|_{\alpha=0}.    \label{Veff2}
\eea

The sum in Eq. (\ref{Veff2}) is calculated with the help of the
$\zeta$-regularization method \cite{zeta}. The main idea is
to treat the power $2\alpha-n$ in the r.h.s. of Eq. (\ref{Veff2})
as a complex parameter, represent the sum in terms of an appropriate
$\zeta$-function assuming that its real part is
large enough to make the sum finite, and then, at the end of calculation,
continue it analytically to the value $\alpha=0$, $\epsilon=0$.
For example, we can write $V_{B}(H)$ in terms
of the Epstein $\zeta$-function \cite{Z-function}, which
for $\mbox{Re} \; \nu > 1$ is defined by
\beq
Z_{1}^{v^{2}} (\nu;w;a) = \sum_{k=-\infty}^{\infty} 
\left[ w (k - a)^{2} + v^{2} \right]^{-\nu}.   \label{Z-def}
\eeq
The bosonic contribution to the effective potential 
is then equal to 
\begin{eqnarray}
V_{B}(H) &=& - \frac{N}{2} 
\frac{ (\pi R^{2}\mu^{2})^{\epsilon}}{R^{4}\pi^{2}} 
 \frac{\partial}{\partial \alpha} \left. \left[ 
  \frac{\Gamma (\alpha-2+\epsilon)}{ 2^{2\alpha} \Gamma (\alpha)}  
  Z_{1}^{0} (\alpha + \epsilon - 2; 1; -u_{B})
 \right] \right|_{\alpha=0} \nonumber  \\
 &=& - \frac{N}{2} \frac{(\pi R^{2}\mu^{2})^{\epsilon}}{R^{4}\pi^{2}} 
 \Gamma (-2+\epsilon) Z_{1}^{0} (-2+\epsilon;1;-u_{B}), 
\label{Veff3}
\end{eqnarray}
where the notation 
\beq
 u_{B} = \frac{1}{2} + \frac{1}{2} R m_{t}(H)  \label{uB}
\eeq
was introduced. In the derivation of formula (\ref{Veff3}) we used the 
relation 
\[
 \left. \frac{\partial}{\partial \alpha} 
 \frac{F(\alpha)}{\Gamma (\alpha)} \right|_{\alpha=0} =  
 F(0) 
\]
valid for any function $F(\alpha)$ analytic at $\alpha = 0$. 

Expanding the r.h.s. of (\ref{Veff3}) in the Laurent series  
in $\epsilon$ at $\epsilon=0$ one gets 
\bea
V_{B}(H) & = & -\frac{N}{2R^{4}\pi^{2}} \left\{
\frac{1}{2\epsilon} Z_{1}^{0} (-2;1;-u_{B}) +
\left[ \frac{1}{2} \left( \ln (\pi \mu^{2} R^{2}) + \frac{3}{2} - 
\gamma_{E} \right) Z_{1}^{0} (-2;1;-u_{B}) \right. \right. \nonumber \\
 & + & \left. \left.  
\frac{1}{2} Z_{1}^{0'}(-2;1;-u_{B}) \right] + {\cal O}(\epsilon) \right\},
\label{Veff5}
\eea
where the prime denotes the derivative of the Epstein function 
with respect to the first argument and $\gamma_{E}$ is the Euler constant.  

The fermionic contribution $V_{F}(H)$ in Eq. (\ref{Veff-dif}) 
is calculated in a similar way. We get 
\bea
V_{F}(H) & = & - \frac{N}{2} 
\frac{(\mu^{2})^{\epsilon}(4\pi)^{\epsilon}}{16\pi^{2}} 
\frac{\partial}{\partial \alpha} \left. 
\frac{ \Gamma (\alpha - 2 + \epsilon)}{ 4^{\alpha} R^{2\alpha} 
\Gamma (\alpha)} \sum_{k=-\infty}^{\infty} 
\frac{1}{\left[ \frac{2k}{R} + m_{t}(H) \right]^{2(\alpha-2+\epsilon)}}
\right|_{\alpha=0}     \label{Veff2-F} \\
 &=& - \frac{N}{2} \frac{(\pi R^{2}\mu^{2})^{\epsilon}}{R^{4}\pi^{2}} 
 \Gamma (-2+\epsilon) Z_{1}^{0} (-2+\epsilon;1;-u_{F}) \nonumber \\
& = & -\frac{N}{2R^{4}\pi^{2}} \left\{
\frac{1}{2\epsilon} Z_{1}^{0} (-2;1;-u_{F}) +
\left[ \frac{1}{2} \left( \ln (\pi \mu^{2} R^{2}) + \frac{3}{2} - 
\gamma_{E} \right) Z_{1}^{0} (-2;1;-u_{F}) \right. \right. \nonumber \\
 & + & \left. \left.  
\frac{1}{2} Z_{1}^{0'}(-2;1;-u_{F}) \right] + {\cal O}(\epsilon) \right\} 
\label{Veff5-F}
\eea
(cf. (\ref{Veff2}), (\ref{Veff3}), (\ref{Veff5})), where 
\beq
 u_{F} = \frac{1}{2} R m_{t}(H).  \label{uF}
\eeq    

For further calculation of the effective potential we will need certain 
properties of the function $Z_{1}^{0}(\nu;1;a)$. We are going to discuss 
them now. For $\mbox{Re} \; \nu > 1$ 
\bea
Z_{1}^{0}(\nu;1;-a) & = &
\sum_{k=-\infty}^{\infty} \frac{1}{[(k-a)^{2}]^{\nu}}
= - a^{-2\nu} + \sum_{k=0}^{\infty} \frac{1}{(k-a)^{2\nu}} +
\sum_{k=0}^{\infty} \frac{1}{(k+a)^{2\nu}} \nonumber \\
& = & \zeta (2\nu,a) + \zeta(2\nu,-a) - a^{-2\nu} \label{Z1},   
\eea
where the Hurwitz $\zeta$-function $\zeta (\nu,a)$ (also called the 
generalized Riemann $\zeta$-function) for 
$\mbox{Re} \; \nu > 1$, $0 < a \leq 1$ is defined by
\beq
  \zeta (\nu, a) = \sum_{k=0}^{\infty} \frac{1}{(k+a)^{\nu}} 
  \label{zeta-def}
\eeq
(see \cite{Ivic}, \cite{Eli}). The function is analytic in $\nu$ in
the whole complex plane except for the simple pole at $\nu =1$.
Two other functions which will be needed for our analysis are 
the Lerch function \cite{Ivic}, 
\beq
\phi (x,a,s) = \sum_{k=0}^{\infty} \frac{e^{2\pi i k x}}{(k+a)^{s}},
\; \; \; \mbox{Re} \;  s > 0, \; \; \; 0 < a \leq 1. \label{Ler-def}
\eeq
and 
\beq
\psi (x,s) = \sum_{k=1}^{\infty} \frac{e^{2\pi i k x}}{k^{s}},
\; \; \; \mbox{Re} \; s > 0. \label{dilog-def}
\eeq
For $s$ being a positive integer $n$ the latter is related to the 
polylogarithm function $Li_{n}(z)$ by $\psi (x,n) = Li_{n}(e^{2\pi i x})$.  

The important functional relation for the Lerch function, which allows the 
analytical continuation of $\zeta (s,a)$ to negative values of 
$s$, is \cite{Ivic}
\beq
\phi (x,a, 1-s) = (2\pi)^{-s} \Gamma (s) \left\{ 
e^{2\pi i (s/4 - ax)} \phi (-a , x, s) + 
e^{2\pi i (-s/4 + a - ax)} \phi (a , 1-x, s) \right\}. 
   \label{Ler-rel}
\eeq 
It is valid for any $s$, $0 < x < 1$ and $0 < a \leq 1$. 
Taking $a=1$ in Eq. (\ref{Ler-rel}) we obtain the following formula: 
\beq
\psi (x,1-s) = (2 \pi)^{-s} \Gamma (s) \left\{ 
e^{i \pi s/2} \zeta (s,x) + 
e^{-i\pi s/2} \zeta (s,1-x) \right\},  
   \label{zeta-rel}
\eeq 
where we have used the relations
\[
\phi (0,a,s) = \zeta(s,a), \; \; \;
\phi (x,1,s) = e^{-2\pi i x} \psi (x,s) 
\]
and the periodicity of the Lerch function in the first argument 
\[
\phi (x+1,a,s) = \phi (x,a,s).
\]
Calculating the real part of the both sides of
Eq. (\ref{zeta-rel}) one gets
\[
\zeta (s,x) + \zeta (s,1-x) = \frac{2 \Gamma (1-s)}{\pi (2 \pi)^{-s}} 
\sin \frac{\pi s}{2} \mbox{Re} \; \psi (x,1-s).
\]
Now we make the change of variabe $s \rightarrow 1-s$, assume that
$\mbox{Re} \; s > 1$ and use representation (\ref{dilog-def}) 
of the $\psi$-function 
in terms of the series. We obtain that 
\beq
\zeta (1-s,x) + \zeta (1-s,1-x) = \frac{2 \Gamma (s)}{\pi (2 \pi)^{s-1}} 
\cos \frac{\pi s}{2} \sum_{k=1}^{\infty}\frac{\cos(2\pi k x)}{k^s}, 
\label{zeta-rel1}
\eeq
The r.h.s. of this relation is valid for
$\mbox{Re} \; s > 0$. Taking into account the property
\beq
\zeta(s, 1+a) = \zeta(s, a) - a^{-s},  \label{zeta-rel2}
\eeq  
one can see from Eq. (\ref{Z1}) that the l.h.s. of (\ref{zeta-rel1})
is equal to $Z_{1}^{0} ((1-s)/2;1;x)$. 
Evaluating it at $s=5-2\epsilon$ and $x=-u$ one gets 
\bea
Z_{1}^{0} (-2+\epsilon;1;-u) & = &
\zeta (-4+2\epsilon,u) + \zeta (-4+2\epsilon,1-u) \nonumber \\
& = & \frac{2}{\pi} \frac{\Gamma (5-2\epsilon)}{(2\pi)^{4-2\epsilon}} 
\sin \pi \epsilon \sum_{k=1}^{\infty}
\frac{\cos(2\pi k u)}{k^{5-2\epsilon}}. \label{Z2}
\eea
{}From this it easy to derive the following formulas: 
\bea
Z_{1}^{0} (-2;1;-u) & = & 0, \label{Z3} \\
Z_{1}^{0'} (-2;1;-u) & = &
\frac{3}{\pi^{4}}\sum_{k=1}^{\infty}\frac{\cos(2\pi k u)}{k^5}.
\label{Z4}
\eea
Substituting these relations into Eqs. (\ref{Veff5}) and (20) we obtain  
that the bosonic and fermionic contributions to 
the 1-loop effective potential in the limit $\epsilon \rightarrow 0$ 
are finite and equal to
\bea
V_{B} (H) & = & - \frac{3N}{4 \pi^{6} R^{4}} \sum_{k=1}^{\infty} 
(-1)^{k} \frac{\cos (\pi k R m_{t} (H))}{k^{5}}, \label{boson}  \\ 
V_{F} (H) & = & -\frac{3N}{4 \pi^{6} R^{4}}
\sum_{k=1}^{\infty}\frac{\cos (\pi k R m_{t} (H))}{k^{5}}.
\label{fermion}
\eea

The renormalization scale $\mu$ does not enter into these 
expressions because its power is multiplied by the same factor as
the pole $1/\epsilon$ (see Eq. (\ref{Veff3})). Substituting results 
(\ref{boson}), (\ref{fermion}) into Eq. (\ref{Veff-dif})
we arrive at the final expression for the 1-loop effective potential
\beq
V_{\mbox{\footnotesize 1-loop}}^{(bare)} (H) = 
V_{\mbox{\footnotesize 1-loop}}^{(R)} (H) = 
\frac{3N}{2\pi^6 R^4}
\sum_{k=0}^{\infty}\frac{\cos\left[(2k+1)\pi R m_t(H)\right]}{(2k+1)^5}.  
\label{final-1}
\eeq
The result in the KK-renormalization scheme is obtained by 
subtracting the $1/\epsilon$ - pole terms. Since they vanish in the  
regularization used here no subtraction is actually needed, and the 
bare result is equal to the renormalized one, as indicated in Eq. 
(\ref{final-1}). It coincides with expression (\ref{final}). 

We have shown that for the regularization used here the bosonic and 
fermionic contributions are separately finite. Therefore the supersymmetric
cancellation is not necessary for getting the finite 1-loop effective 
potential. Clearly the main difference between our approach and the 
previous ones, namely the 
KK-regularization~\cite{ADPQ, delgado, barbieri1, arkani}
and the proper time regularization~\cite{benakli2, tatsuo}, is in the
way the ultraviolet divergencies reveal themselves in the 
regularized theory.

Let us analyze its ultraviolate properties in more detail.
As it was already mentioned in the Introduction, the pole $1/\epsilon$
is multiplied by the sum of three terms: the term with $k=0$ (zero mode
contribution), the sum of contributions of modes with $k > 0$ and the
sum of contributions of modes with $k < 0$ (see Eqs. (\ref{Veff5}), 
(\ref{Veff5-F}), (\ref{Z1})).
For example, in the case of the fermionic contribution they
correspond to top quark contribution, to the contribution of non-zero
KK modes $X^{+}=\psi_{t}^+$ and to the contribution of
non-zero KK modes $X^{-}=\psi_{t}^{-}$ respectively. The
divergent part $V_F^{(div)}(H)$
can be written as:
\begin{equation}
V_F^{(div)}(H) =
-\frac{N}{64 \pi^2 \epsilon}\left(m_t(H)^4 +
\sum_{k=1}^{\infty}\left(\frac{2k}{R} + m_t(H)
\right )^4 + \sum_{k=1}^{\infty}\left(-\frac{2k}{R} +
m_t(H)\right )^4 \right)
\label{vinfinito}
\end{equation}
where $m_t(H)$ is the mass of the top quark (see Eq. (\ref{mtop})).
Here the infinite sums are understood as being regularized by
the $\zeta$-regularization, namely 
\bea
\sum_{k=1}^{\infty} \left( \frac{2k}{R} + m_{t}(H) \right)^{4} & = &
\frac{16}{R^{4}} \zeta(-4,u_{F}) - m_{t}^{4}(H), \nonumber \\
\sum_{k=1}^{\infty} \left( - \frac{2k}{R} + m_{t}(H) \right)^{4} & = &
\frac{16}{R^{4}} \zeta(-4,-u_{F}) - m_{t}^{4}(H). \nonumber
\eea
We have shown that the three terms in Eq. (\ref{vinfinito})
cancel each other and
\bea
V_F^{(div)}(H) & = & - \frac{N}{64 \pi^{2} \epsilon} 
\left[ - m_{t}(H)^{4} + \frac{16}{R^{4}} \zeta (-4,u_{F}) + 
\frac{16}{R^{4}} \zeta (-4,-u_{F}) \right] \nonumber \\
& = & - \frac{N}{2 R^{4} \pi^{2}} \frac{1}{2 \epsilon} 
\left[ \zeta (-4,u_{F}) + \zeta (-4,1-u_{F}) \right] = 0. \nonumber
\eea
The cancellation occurs because the KK modes, contributing to the 
1-loop effective potential, combine into a single tower whose 
spectrum is given by (\ref{mf}) with $k$ running
from $k=-\infty$ to $k=\infty$.
In general such cancellation does not occur in models
where higher KK modes are truncated.

A generalization of the Epstein $\zeta$-function to
the case of $d$ extra dimensions is given by
\[
Z_{d}^{v^{2}}(\nu;w_{1}, \ldots , w_{d}; u_{1}, \ldots , u_{d}) =
\sum_{k_{1}, \ldots , k_{d}=-\infty}^{\infty}
\left[ w_{1} (k_{1}-u_{1})^{2} + \ldots +
w_{d} (k_{d}-u_{d})^{2} + v^{2}\right]^{-\nu}.
\]
The property of this function, which is important for us, is~\cite{Z-function}
(see also \cite{EliKub})
\beq
Z_d^{v^2}(-2;w_1, \ldots ,w_d;u_1, \ldots ,u_d) =\left\{
                   \begin{array}{cl}
             0, &  \mbox{for }\ d\  \mbox{odd},\\
             \frac{\dsp 2(-1)^{\frac d 2}\pi^{\frac d
2}v^{d+4}}{\dsp \sqrt{w_1...w_d}\left(\frac d 2 +2 \right)!},   &
 \mbox{for}\  d\     \mbox{even}.
        \end{array}\right.                  \label{sta3}
\eeq

One important observation is that, in fact, the 1-loop effective potential
in a 5D generalization of the SM is finite for any particle independently of
the concrete form of the spectrum. For example, 
KK modes in the gauge sector have the masses given by~\cite{delgado}
\begin{equation}
m_{G,k}^2 = \frac{(k + q_G)^2}{R^2} + f_G (H)^2 , 
\label{gaugemass}
\end{equation} 
where the lower index $G$ stands for the gauge bosons and the gauginos,
$f_G(H)^2$ is some function of the Higgs field which is proportional
to the gauge coupling, and $q_G$ is a constant
whose value is 0 (1/2) for the gauge bosons (gauginos) in the model
of Ref.~\cite{barbieri1} and is arbitrary in models of
Ref.~\cite{delgado, delgado2}. The contribution of the gauge sector in the
Landau gauge to the divergent part of the effective potential
is equal to
\begin{equation} 
V_G^{(div)} (H) = -\frac{3}{2 R^4 \pi^2 \epsilon} Z_1^{v_{G}^2}(-2;1;-q_G), 
\label{vgauge}
\end{equation}
where $v_{G}^{2} = R^{2} f_{G}(H)^{2}$. From formula (\ref{sta3}) we get
$Z_1^{f_G(H)^2}(-2;1;-q_G) = 0$ and, hence, the divergent part
of the effective potential is zero both for the gauge boson and for the
gaugino separately.

According to Eq. (\ref{sta3})
the contribution to the 1-loop effective potential of an individual field
is in general not finite if the number of extra dimensions is even.
In this case the finiteness of the theory relies on
the SUSY cancellation between bosons and fermions.
It can be easily seen from Eq. (\ref{sta3}) that, since
$Z_n^{v^2}(-2;w_n;u_n)$ is independent of the soft masses,
entering into $u_i$, the divergences cancel each other by the SUSY mechanism
and the 1-loop effective potential is finite provided the bosons
and fermions have the same Higgs field dependent part 
(analog of $f_{G}(H)$ in Eq. (\ref{gaugemass})) in their spectrum.
Also, using property (\ref{sta3}) we conclude that as long as
the spectra are of the form
\begin{equation}
m_{k_{1}, \ldots , k_{d}}^{2} \propto \frac{1}{R^{2}}
\left[ w_{1} (k_{1}-u_{1})^{2} + \ldots +
w_{d} (k_{d}-u_{d})^{2} \right],
\label{linealmass}
\end{equation}
the 1-loop effective potential is finite without supersymmetry 
in any number of extra dimensions. In particular, the linear spectrum 
of the model considered above, given by Eqs. (\ref{mb}), (\ref{mf}), 
is of this type. In Ref.~\cite{benakli2}, using the proper time regularization,
the effective potential in a model with an arbitrary number of
extra dimensions and with the spectrum of type
(\ref{linealmass}) was calculated.
It was shown there that without SUSY the effective potential
had a divergent term associated with the cosmological constant.
In our approach the field-independent divergence vanishes automatically.

%%%%%%%%%%%%%%%%%%%%%%%%%%%%%%%%%%%%%%% Sect. 4 R -> 0 LIMIT 

\section{$R \rightarrow 0$ limit}

A natural question to pose is whether the 1-loop effective 
potential of the four-dimensional SM is recovered in the 
limit $R \rightarrow 0$. One can calculate the limit of the final 
expressions (\ref{Z2}) - (\ref{final-1}) taking into account 
that 
\[
R m_{t}(H) \sim R(y_{t} H) \rightarrow 0, \; \; \; 
u_{B} \sim \frac{1}{2} +  \frac{1}{2} R(y_{t} H) \rightarrow \frac{1}{2} \; \; \; 
u_{F} \sim  \frac{1}{2} R(y_{t} H) \rightarrow 0 
\]
when $R \rightarrow 0$ and the properties 
\[
\zeta \left( s, \frac{1}{2} \right) = (2^{s}-1) \zeta (s), \; \; \; 
\zeta (s,1) = \zeta (s), 
\]
where $\zeta (s)$ is the Riemann $\zeta$-function (see for example 
\cite{Ivic}, \cite{zetaR}). Using Eqs. (\ref{zeta-rel2}), (\ref{Z2}) 
one gets 
\bea
Z_{1}^{0}(\nu;1;-u_{B}) & = & 
\zeta \left (2\nu, \frac{1}{2} + \frac{1}{2} R m_{t}\right) + 
\zeta \left (2\nu, \frac{1}{2} - \frac{1}{2} R m_{t}\right) 
\rightarrow 2 (2^{2\nu} -1) \zeta (2\nu), \label{Z3-1}  \\
Z_{1}^{0}(\nu;1;-u_{F}) & = & 
\zeta \left (2\nu, 1 + \frac{1}{2} R m_{t}\right) + 
\zeta \left (2\nu, 1 - \frac{1}{2} R m_{t}\right) + 
\left( \frac{1}{2} R m_{t} \right)^{-2\nu} \nonumber \\ 
&\rightarrow & 2 \zeta (2\nu) + \left( \frac{1}{2} R m_{t} \right)^{-2\nu}. 
\label{Z4-1}
\eea 
Substituting $\nu = -2 -\epsilon$ into (\ref{Z3-1}) one obtains the 
following formulas necessary for the calculation of the bosonic 
contribution in the limit $R \rightarrow 0$:
\bea
Z_{1}^{0} \left(-2;1;-\frac{1}{2} \right) & = & - \frac{15}{8} \zeta (-4) = 0, 
\nonumber \\
Z_{1}^{0'} \left(-2;1;-\frac{1}{2} \right) & = & 
\frac{1}{4} \ln 2 \zeta (-4) - \frac{15}{4} \zeta' (-4) = - \frac{3}{\pi^{4}} 
\frac{15}{16} \zeta (5).  \label{Z5}  
\eea
Here we used the relations $\zeta (-4) = 0$ and 
$\zeta'(-4) = 3 \zeta (5)/(4\pi^{4})$ \cite{Ivic}, \cite{zetaR}. They 
follow from a functional equation for the Riemann $\zeta$-function  
which relates $\zeta (s)$ to $\zeta (1-s)$\footnote{This equation is actually 
a particular case of Eq. (\ref{Ler-rel}) for $a=1$, $x=1/2$.}. 

The fermionic contribution in the limit $R \rightarrow 0$ is given by the Eq. (\ref{Z4-1}) at 
$\nu = -2 + \epsilon$  
\bea
Z_{1}^{0} \left(-2;1;0 \right) & = & 2\zeta (-4) = 0, 
\nonumber \\
Z_{1}^{0'} \left(-2;1;0 \right) & \sim & 
4 \zeta' (-4) - \left. \left( \frac{1}{2} R m_{t} \right)^{4} 
\ln \frac{R^{2} m_{t}^{2}}{4} \right|_{R \rightarrow 0}  = 
\frac{3}{\pi^{4}} \zeta (5). \label{Z6}  
\eea

Substituting (\ref{Z5}) and (\ref{Z6}) into Eqs. (\ref{Veff5}) and (\ref{Veff5-F}) 
we obtain that in the limit $R \rightarrow 0$ the potential is divergent in 
$R$ with the leading term being 
\beq
V_{\mbox{\footnotesize 1-loop}}^{(bare)} (H) = V_B(H) - V_F(H) = 
V_{\mbox{\footnotesize 1-loop}}^{(R)} (H) \sim \frac{3N}{4 R^{4} \pi^{6}} 
\zeta (5).   \label{R0-V1}
\eeq
As one could expect, the 1-loop contribution does not contain 
$1/\epsilon$-pole. Note also that in the calculation of 
$V_{\mbox{\footnotesize 1-loop}}^{(R)} (H)$ it was basically assumed that 
$R H \ll 1$, i.e. Eq. (\ref{R0-V1}) can be considered as the limit of weak 
field, $H \ll 1/R$. 

The limit $R \rightarrow 0$ can also be studied using Eqs. (\ref{Veff2}), 
(\ref{Veff2-F}) as the starting point. If we take the limit formally before 
summing over $k$, then the only non-vanishing contribution is 
the one of the fermionic mode with $k=0$ in Eq. (\ref{Veff2-F}). By a 
straightforward calculation we arrive at the result 
\bea
V_{\mbox{\footnotesize 1-loop}}^{(bare)} (H) & = & - V_{F}(H) = 
- \frac{N}{2} \frac{\mu^{2\epsilon} (4\pi)^{\epsilon}}{16 \pi^{2}}
\Gamma (-2+\epsilon) (y_{t}H)^{4-2\epsilon} + {\cal O}(R) \nonumber \\
& = &  \frac{N}{64 \pi^{2}} (y_{t}H)^{4} \left[ \frac{1}{\epsilon} 
+ \left( \frac{3}{2} - \gamma_{E} - \ln \frac{y_{t}^{2} H^{2}}{4\pi \mu^{2}} 
\right) + {\cal O}(\epsilon,R) \right]. \label{R0-V2}
\eea
This is the known expression for the 1-loop contribution to the 
effective potential in the SM. 

In four dimensions the KK-renormalization scheme reduces to the $\overline{\mbox{MS}}$ scheme. 
Subtracting ($1/\epsilon - \gamma_{E} + \log 4\pi$) in Eq. (\ref{R0-V2}) we get 
\beq
V_{\mbox{\footnotesize 1-loop}}^{(R)} (H) 
= - \frac{N}{64 \pi^{2}} (y_{t}H)^{4} 
\left( \ln \frac{y_{t}^{2} H^{2}}{ \mu^{2}} - \frac{3}{2} \right) + {\cal O}(R). \label{R0-V3}
\eeq

Results (\ref{R0-V1}) and (\ref{R0-V3}) are obviously different. The 
reason is that Eq. (\ref{R0-V1}) was calculated by first summing over 
the KK modes by means of the $\zeta$-regularization and then taking the 
limit $R H \rightarrow 0$. The summation was performed over the 
complete infinite KK tower of modes, including heavy modes with 
the masses $m_{k}^{(B)}, m_{k}^{(F)} \sim k/R \gg 1$. 

Contrary to this, Eq. (\ref{R0-V3}) was derived by first taking the limit 
$R \rightarrow 0$ in Eqs. (\ref{Veff2}), (\ref{Veff2-F}). After this 
the sums over KK modes reduce just to the term with $k=0$ in $V_{F}(H)$. 
The non-zero KK modes, i.e. all modes in (\ref{mb}) and all modes with 
$k \neq 0$ in (\ref{mf}), decouple and the four-dimensional results 
(\ref{R0-V2}), (\ref{R0-V3}) emerge.   

The conclusion is that the summation over the KK modes by means of the 
$\zeta$-regularization and the limit $R \rightarrow 0$ do not commute 
within the KK-renormalization scheme. Analyzing the limit of the 
5D expression for the renormalized 1-loop effective potential, 
Eq. (\ref{R0-V1}), we see that the contribution of 
KK modes does not decouple. 
Therefore, the KK-renormalization scheme is {\it not} a scheme with explicit 
decoupling of heavy modes \cite{Col}. We would like to note that 
a renormalization scheme with explicit decoupling of KK modes, 
based on a certain momentum subtraction procedure, was developed and used 
for calculations in multidimensional theories \cite{EliKub}, \cite{KOCS}. 

%%%%%%%%%%%%%%%%%%%%%%%%%%%%%%%%%%%%%%%%% Sect. 5 Conclusions

\section{Conclusions}

We have discussed a technique for calculation of the 1-loop
effective potential and considered the problem of its finiteness
in 5D SUSY theories. As a simple example the model proposed in
Ref.~\cite{barbieri1} was used.

As we mentioned in the Introduction, 
recently there have been some discussion in the literature
regarded a controversy in the calculation the effective potential~
\cite{nieles, kim}. In particular it was pointed out that for each KK mode
the 1-loop integral, contributing to the effective potential, is
ultraviolet divergent and does not commute with the summation, hence
it is unclear whether the KK-regularization is an appropriate procedure.
It was also argued that KK modes above some ``natural'' scale of the
theory can not contribute to the effective potential and, therefore,
a cutoff of the sum over KK modes should be introduced.

The KK-renormalization  proposed in the present paper
overcomes successfully the first problem by combining the 
dimensional regularization to evaluate the integrals and
the zeta-regularization to calculate the infinte sums.
Since both regularizations are based on analytic continuation of the
integrand and the summand, no difficulty arises in interchanging
the summation and integration. In general, with such regularization the
result of calculation may contain a pole $1/\epsilon$ coming from the pole
of the generalized Riemann $\zeta$-function (see Eq. (\ref{Z1}) and the
remark below Eq. (\ref{zeta-def})). The KK-renormalization
procedure consists in subtraction of the pole terms, and is 
similar to the ($\overline{\mbox{MS}}$) scheme.
When the final result is finite, as in the
model considered in the article, obviously no subtraction is needed.
Our technique for calculation of the 1-loop effective potential
reproduces the finite result obtained by using regularizations 
respecting the analytical symmetry of the
contributions \cite{barbieri1}.

The KK-renormalization introduces no cutoff, neither in the integrals over 
four dimensional 
momentum, nor in the sums over KK modes. In this sense the sums over $k$ 
are treated in the same fashion as the integrals,
thus making the whole procedure consistent. Recall that $k$, the number 
of a KK mode, plays the role of the (discrete) momentum in the 
fifth direction (in terms of $R^{-1}$). 

In the model considered here both the KK-regularization and
our approach give the finite 1-loop effective potential.
However, in the KK-regularization the bosonic and fermionic contributions
are separately divergent and cancel out due to the SUSY mechanism.
Instead, in our approach the bosonic and fermionic contributions to
the effective potential are separately finite, and the supersymmetry is not
necessary for the finiteness of the theory at 1-loop. In Sect. 3 
we have also shown that for the torus compactification the 
finiteness at 1-loop order is rather generic and independent of 
concrete details of the spectrum in the theory.

As we have seen, what is really important for the finiteness
of the effective potential at 1-loop order is that the spectrum of
KK modes effectively corresponds to compactification to the circle $S^{1}$,
i.e. it is given by formulas of type (\ref{mb}), (\ref{mf})
with $k$ running from $k=-\infty$ to $k=\infty$. This assures that
the contributions to the divergent part from the zero mode, positive modes
and negative modes cancell each other exactly. 

We also studied the 1-loop effective potential in the limit 
$R H \rightarrow 0$, i.e. when either the Higgs background field 
$H \ll 1/R$, or the space of extra dimension collapses to a point. 
We showed that within the KK-renormalization scheme the summation over the 
KK modes and the limit $R \rightarrow 0$ do not commute, i.e. the 
KK-renormalization scheme is not a scheme with explicit decoupling of heavy 
KK modes.

\bigskip
{ \large \bf Acknowledgments}
\medskip

We would like to thank M. Quir\'os for useful and illuminating comments.
V.D.C would like to thank PPARC for a Research Associateship.
Yu.K. acknowledges financial support from the Royal Society
Short Term Visitor grant ref. RCM/ExAgr/hostacct and
the Russian Foundation for Basic Research (grant 00-02-17679).


\begin{thebibliography}{99}

\bibitem{antoniadis}
I. Antoniadis, {\it Phys. Lett.} {\bf B246} (1990) 377. \\
I. Antoniadis, C. Mu\~noz and M. Quir\'os, {\it Nucl. Phys.}
{\bf B397} (1993) 515. \\
I. Antoniadis and K. Benakli, {\it Phys. Lett.} {\bf B326} (1994) 69. \\
I. Antoniadis, K. Benakli and M. Quir\'os, {\it Phys. Lett.}
{\bf B331} (1994) 313. \\
K. Benakli, {\it Phys. Lett.} {\bf B386} (1996) 106. \\
I. Antoniadis and M. Quir\'os, {\it Phys. Lett.} {\bf B392} (1997) 61. \\
E.A. Mirabelli and M. Peskin, {\it Phys. Rev.} {\bf D58} (1998) 65002. \\
A. Pomarol and M. Quir\'os, {\it Phys. Lett.} {\bf B438} (1998) 255.
%
\bibitem{ADPQ}
I. Antoniadis, S. Dimopoulos, A. Pomarol, and M. Quir\'os,
{\it Nucl. Phys.} {\bf B544} (1999) 503.
%
\bibitem{delgado} A. Delgado, A. Pomarol and M. Quir\'os,
{\em Phys. Rev.} {\bf D60} (1999) 095008.
%
\bibitem{barbieri1} R. Barbieri, L.J. Hall and Y. Nomura,
{\it Phys. Rev} {\bf D63} (2001) 105007.
%
\bibitem{arkani}
N. Arkani-Hamed, L.J. Hall, Y. Nomura, D. Smith and N. Weiner,
{\it Nucl. Phys.} {\bf B605} (2001) 81. 
%
\bibitem{delgado2}  
A. Delgado and M. Quir\'os, {\it Nucl. Phys.} {\bf B607} (2001) 99. 
%
\bibitem{barbieri2}  
R. Barbieri, L.J. Hall and Y. Nomura, {\rm hep-th/0107004}.
%
\bibitem{antoniadis1}  
I. Antoniadis, K. Benakli and M. Quir\'os, {\it Nucl. Phys.} 
{\bf B583} (2000) 35. \\
I. Antoniadis and K. Benakli, {\it Int. J. Mod. Phys. A} {\bf 15} 
(2000) 4237. \\ 
I. Antoniadis, {\rm hep-th/0102202}. 
%
\bibitem{mariano} 
A. Delgado, G. von Gersdorff and M. Quir\'os, {\it Nucl. Phys.} {\bf B613} 
(2001) 49. 
%
\bibitem{masiero}
A. Masiero, C.A. Scrucca, M. Serone and L. Silvestrini, 
{\it Phys. Rev. Lett.} {\bf 87} (2001) 251601.
%
\bibitem{nieles}
D.M. Ghilencea and H.P. Nilles, {\it Phys. Lett.} {\bf B507} (2001) 327.
%
\bibitem{kim}
H.D. Kim, {\rm hep-ph/0106072}.
%
\bibitem{gero}
A. Delgado, G. von Gersdorff, P. John and M. Quir\'os, {\it Phys. Lett.} 
{\bf B517} (2001) 445. \\ 
R. Contino and L. Pilo, {\it Phys. Lett.} {\bf B523} (2001) 347.
%
\bibitem{zeta}
D. Ray and I. Singer, {\em Adv. Math.} {\bf 7} (1971) 145. \\ 
J.S. Dowker and R. Critchley, {\em Phys. Rev.} {\bf D13} (1976) 3224. \\ 
S. Hawking, {\em Commun. Math. Phys.} {\bf 55} (1977) 133. 
%
\bibitem{zeta-KK}
T. Appelquist and A. Chodos, {\em Phys. Rev. Lett.} {\bf 50} (1983) 141;  
{\em Phys. Rev.} {\bf D28} (1983) 772. \\
I. Antoniadis and M. Quir\'os, {\em Nucl. Phys.} {\bf B505} (1997) 109. 
% 
\bibitem{EliKub}
E. Elizalde, K. Kirsten and Yu. Kubyshin, {\em Z.Phys.} {\bf C70} (1996) 159. 
%
\bibitem{corneliu}
C.~Sochichiu, {\it Phys. Lett.} {\bf B463} (1999) 27. \\
W.~D.~Goldberger and I.~Z.~Rothstein, {\it Phys. Lett.} 
{\bf B491} (2000) 339. \\ 
E.~Ponton and E.~Poppitz, {\it JHEP} {\bf 0106} (2001) 019.
%
\bibitem{nomura}
Y. Nomura, {\rm hep-ph/0105113}. \\ 
N. Weiner, {\rm hep-ph/0106021}.
%
\bibitem{benakli2}
I. Antoniadis, K. Benakli and M. Quir\'os, {\rm hep-th/0108005}.
%
\bibitem{Z-function}
K. Kirsten, {\em J. Phys.} {\bf A26} (1993) 2421. \\ 
E. Elizalde and K. Kirsten, {\em J. Math. Phys.} {\bf 35} (1994) 1260. \\
E. Elizalde, S.D. Odintsov, A. Romeo, A.A. Bytsenko and S. Zerbini, 
``Zeta Regularization Techniques with Applications'', 
{\it World Scientific}, Singapore (1994). \\
E. Elizalde, "Ten Physical Applications of Spectral Zeta Functions", 
{\em Lecture Notes in Physics} M35, 1, Springer (1995).  
%
\bibitem{Ivic}
A. Ivi\'c,``The Riemann zeta function: the theory of the Riemann 
zeta-function with applications'', {\it A Wiley - Interscience Publ.}, 
N.Y., 1985. 
%
\bibitem{Eli}
E. Elizalde, {\em J. Phys.} {\bf A22} (1989) 931; E. Elizalde,
{\em J. Math. Phys.} {\bf 31} (1990) 170.
%
\bibitem{tatsuo} 
T. Kobayashi and H. Terao, {\rm hep-ph/0108072}.
%
\bibitem{zetaR}
H.M. Edwards, "Riemann's Zeta Function", {\em Acad. Press.}, N.Y., 1974. 
%
\bibitem{Col}
J.C. Collins, "Renormalization", {\em Cambridge Univ. Press.}, 1985. 
%
\bibitem{KOCS}
Yu.A. Kubyshin, D. O'Connor and C.R. Stephens, {\em Class. Quant.
Grav.} {\bf 10} (1993) 2519; "Decoupling of heavy masses in the 
Kaluza-Klein approach". In: "Quarks '92",  Proc. of the Intern. 
Seminar. Eds. D.Yu. Grigoriev et al. {\em World Scient.} 1993, 359.
%

\end{thebibliography}
\end{document}